\documentstyle[12pt]{article}

\def\ar{\rightarrow}
\def\bib{\bibitem}
\def\intx{\int\! d^{\sl 4}x}
\def\intX{\int\! d^{\sl 4}X\,}

\def\intP{\int\! \frac{d^{\sl 4}P}{(2{\pi})^4}\,}

\def\lar{\longrightarrow}
\def\pa{\partial}
\def\rvec{\!\!\!\!^{^\rightarrow}}

\def\Tr{\,\mbox{Tr}\,}

\def\al{\alpha}
\def\be{\beta}
\def\ga{\gamma}
\def\de{\delta}
\def\ep{\varepsilon}

\def\la{\lambda}

\def\si{\sigma}
\def\om{\omega}

\def\La{{\it\Lambda}}
\def\Om{{\it\Omega}}

\def\Pit{\!{\it\Pi}}
\def\Pitch{\!{\tilde{\hat{\it\Pi}^*}}}
\def\Pitth{\!{\tilde{\hat{\it\Pi}}}}

\def\Th{{\it\Theta}}

\def\beq{\begin{equation}}
\def\eeq{\end{equation}}
\def\bed{\begin{displaymath}}
\def\eed{\end{displaymath}}
\def\beqq{\begin{eqnarray}}
\def\eeqq{\end{eqnarray}}
\def\bedd{\begin{eqnarray*}}
\def\eedd{\end{eqnarray*}}

\begin{document}

\centerline{\normalsize\bf I - CONSERVATION OF GRAVITATIONAL ENERGY-MOMENTUM} \centerline{\normalsize\bf AND INNER DIFFEOMORPHISM GROUP GAUGE INVARIANCE}

\vspace*{0.9cm}
\centerline{\footnotesize C. WIESENDANGER}
\baselineskip=12pt
\centerline{\footnotesize\it Aurorastr. 24, CH-8032 Zurich}
\centerline{\footnotesize E-mail: christian.wiesendanger@ubs.com}

\vspace*{0.9cm}
\baselineskip=13pt
\abstract{Viewing gravitational energy momentum $p_G^\mu$ as equal by observation, but different in essence from inertial energy-momentum $p_I^\mu$ requires two different symmetries to account for their independent conservations - spacetime and inner translation invariance. Gauging the latter a generalization of non-Abelian gauge theories of compact Lie groups is developed resulting in the gauge theory of the non-compact group of volume-preserving diffeomorphisms of an inner Minkowski space ${\bf M\/}^{\sl 4}$. As usual the gauging requires the introduction of a covariant derivative, a gauge field and a field strength operator. An invariant and minimal gauge field Lagrangian is derived. The classical field dynamics and the conservation laws for the new gauge theory are developed. Finally, the theory's Hamiltonian in the axial gauge is expressed by two times six unconstrained independent canonical variables obeying the usual Poisson brackets and the positivity of the Hamiltonian is related to a condition on the support of the gauge fields.}

\normalsize\baselineskip=15pt

\section{Introduction}
Field theory provides a powerful way to represent fundamental conservation laws of Nature in a mathematically consistent framework through Noether's theorem which relates any global invariance of the underlying field theory under a continous symmetry group to a number of conserved currents and charges. Conservation of electric charge and $U(1)$-invariance, conservation of Color and $SU(3)$-invariance or conservation of inertial energy-momentum and translation invariance in spacetime are but three key examples.

Moreover, through the gauge principle the field theory framework allows to construct new fields together with their dynamics and through minimal coupling it allows to fix the coupling to other fields obeying a given global symmetry in a way which extends it to a local invariance of the thus completed field theory \cite{lor}. The new fields transmit the physical interactions between the various minimally coupled fields (and between themselves in all cases with a non-Abelian underlying symmetry). The dynamics of the Standard Model (SM) has been modelled along this way starting with Electrodynamics, extending it to electro-weak interactions and finally adding Chromodynamics which models the strong interaction \cite{stw1,stw2}. And General Relativity (GR) can be constructed along this way as well \cite{chw1}.

There is, however, a crucial difference between the SM and GR when attempting to quantize the respective classical gauge field theories. The quantized SM is a perfectly consistent quantum field theory (QFT) related to its renormalizability - at least at the perturbation theory level. Any attempt to consistently quantize GR or extensions thereof have failed so far already at the perturbation theory level due to the intrinsic non-renormalizability of the theory \cite{car,clk}.

Whereas in the SM spacetime and its Minkowskian geometry are an A Priori which serves as the arena within which the dynamics of the various matter and gauge fields unfolds GR declares spacetime itself a dynamical element whose geometry evolves alongside the changing energy-momentum distribution of the matter and gauge fields present. Not only does this dynamization of spacetime render GR non-renormalizable, but it also destroys basic concepts such as the energy-momentum density of the gravitational field or the relation between a quantized field and its corresponding particle which both rely on global translation invariance in spacetime and are crucial for the physical interpretation of a QFT \cite{car,clk}.

To map out an alternative route to a viable field theory of gravitation let us go back to the very fundament - namely the two experimental observations that (1) the inertial and gravitational masses of a physical body are numerically equal, $m_I = m_G$, and (2) the inertial energy-momentum of a closed physical system is conserved, $p_I^\mu = conserved$. Taking both together we then can write in the rest frame of the body
\beq \label{1}
p_I^\mu = (m_I,\underline{0}) =^{\!\!\!\!\!\!\! ^{ (1)}} (m_G,\underline{0}) = p_G^\mu,
\eeq
where we have tentatively introduced the gravitational energy-momentum $p_G^\mu$ which we keep as an entity a priorily different from $p_I^\mu$. Note that $p_G^\mu$ is conserved due to Eqn.(\ref{1}).

In GR $m_I = m_G$ is interpreted as an essential identity which leads to the aforementioned geometrical description of gravitation.

In this paper we propose to follow a different route and {\it investigate the consequences of viewing $m_I$ and $m_G$ or $p_I^\mu$ and $p_G^\mu$ as different by their very natures} - the prevailing view before Einstein which comes at the price of accepting the observed numerical equality $m_I = m_G$ as accidential.

Both $p_I^\mu$ and $p_G^\mu$ are four-vectors then which are conserved, but through two different mechanisms. Obviously the conservation of $p_I^\mu$ is related to translation invariance in spacetime. Let us use Noether's theorem to separately derive the conservation of a new four-vector in a field theoretical framework relating it to a continous symmetry of the theory which we will call inner translation invariance. That four-vector is then interpreted as the gravitational energy-momentum $p_G^\mu$. 

This will be the first step in developing an alternative route to describe gravity. The second will be to gauge the inner translation group and to develop the gauge field theory of inner diffeomorphisms technically leveraging earlier work on generalizing Yang-Mills theories to gauge groups with infinitely many degrees of freedom \cite{chwA,chwB}. The resulting Lagrangian and Hamiltonian dynamics are the basis to interpret the theory as a theory of gravitation \cite{chw2} and to show its renormalizability at the QFT level \cite{chw3} in two forthcoming papers.

The notations and conventions used follow closely to those of Steven Weinberg in his classic account on the quantum theory of fields \cite{stw1, stw2}. They are presented in the Appendix.

\section{Global Diffeomorphism Invariance and Conservation of Gravitational Energy-Momentum}
In this section we introduce the concept of global diffeomorphism invariance in inner space for a generic field theory in order to generate a new conserved four-vector through Noether's theorem which will serve as gravitational energy-momentum. 

Let us start with a four-dimensional real vector space
${\bf R}^{\sl 4}$ with elements labelled $X^\al$ without a metric structure at this point which we will call inner space in the following. Volume-preserving diffeomorphisms 
\beq \label{2}
X^\al\lar X'^\be = X'^\be(X^\al),\:\:\al,\be=\sl{0,1,2,3}
\eeq
act as a group ${\overline{DIFF}}\,{\bf R}^{\sl 4}$ under composition on this space. $X'^\be (X)$ denotes an invertible and differentiable coordinate transformation of ${\bf R\/}^{\sl{4}}$ with unimodular Jacobian
\beq \label{3}
\det\left(\frac{\pa X'^\be (X)}{\pa X^\al}\right) = 1.
\eeq
The restriction to volume-preserving transformations will automatically ensure global gauge invariance of the theories we look at in the sequel and will prove crucial for the consistency of our approach.

To represent this group in field space we have to add additional degrees of freedom in complete analogy to the Yang-Mills case where fields become vectors on which representations of a finite-dimensional symmetry group act.

Hence we consider fields $\psi(x,X)$ defined on the product of the four-dimensional Minkowski spacetime ${\bf M}^{\sl 4}$ and the four-dimensional inner space ${\bf R}^{\sl 4}$ introduced above. The fields $\psi(x,X)$ are assumed to be infinitely differentiable in both $x$ and $X$ and to vanish at infinity. They form a linear space endowed with the scalar product
\beq \label{4}
\langle \psi \!\mid\! \chi \rangle \equiv \intx\intX \La^{-4} \psi^\dagger (x,X) \cdot\chi(x,X),
\eeq
where we introduce a parameter $\La$ of dimension length, $[\La]=[X]$, so as to define a dimensionless scalar product. $\La$ will play an important role in the definition of the gauge field action later.

Note that the fields $\psi(x,X)$ might live in non-trivial representation spaces of both the Lorentz group with spin $s\not= 0$ and of other inner symmetry groups such as $SU(N)$. All these scalar, spinor and gauge vector fields - apart from the gauge field related to diffeomorphism invariance to be introduced below - are called "matter" fields in the following. These representations factorize w.r.t the diffeomorphism group representations we introduce below which is consistent with the Coleman-Mandula theorem.

Let us assume in the sequel that the dynamics of the field $\psi(x,X)$ is specified by a Lagrangian of the form 
\beq \label{5}
L_M(\psi, \pa_\mu \psi) = \intX\La^{-4} {\cal L}_M (\psi(x,X), \pa_\mu \psi(x,X))
\eeq
with a real Lagrangian density ${\cal L}_M$. The integration measure in inner space comes along with a factor of
$\La^{-4}$ to keep inner integrals dimensionless. The subscript $_M$ denotes generic fermionic and bosonic matter in this context

Turning to the transformation behaviour of the Lagrangian Eqn.(\ref{5}) under infinitesimal diffeomorphism transformations we start with the passive representation of ${\overline{DIFF}}\,{\bf R}^{\sl 4}$ in field space for infinitesimal transformations $X'^\al (X) = X^\al + {\cal E}^\al (X)$
\beqq \label{6}
x^\mu &\lar& x'^\mu = x^\mu, \quad
X^\al \lar X'^\al = X^\al, \\
\psi (x,X) &\lar& \psi'(x,X) = \psi(x,X) -\, {\cal E}^\al (X)\cdot \nabla_\al\,\psi(x,X) \nonumber
\eeqq
transforming the fields only.

The unimodularity condition Eqn.(\ref{3}) translates into the infinitesimal gauge parameter ${\cal E}^\al$ being divergence-free
\beq \label{7}
\nabla_\al {\cal E}^\al (X) = 0. \nonumber
\eeq
Note the crucial fact that the algebra ${\overline{\bf diff}}\,{\bf R}^{\sl 4}$ of the divergence-free ${\cal E}$s closes under commutation. For $\nabla_\al {\cal E}^\al = \nabla_\be {\cal F}^\be = 0$ we have
\beq \label{8} 
\left[{\cal E}^\al \cdot \nabla_\al, {\cal F}^\be \cdot \nabla_\be \right] = \left( {\cal E}^\al \cdot \nabla_\al {\cal F}^\be 
- {\cal F}^\al \cdot \nabla_\al {\cal E}^\be \right) \nabla_\be
\eeq
with
\beq \label{9} 
\nabla_\be \left({\cal E}^\al \cdot \nabla_\al {\cal F}^\be 
- {\cal F}^\al \cdot \nabla_\al {\cal E}^\be \right) = 0
\eeq
as required by the finite transformations ${\overline{DIFF}}\,{\bf R}^{\sl 4}$ forming a group under composition.

As a result we can write infinitesimal transformations in field space 
\beq \label{10}
U_{\cal E} (X) \equiv {\bf 1} - {\cal E}(X),\:\: {\cal E}(X)= {\cal E}^\al (X)\cdot \nabla_\al
\eeq
as anti-unitary operators w.r.t. the scalar product Eqn.(\ref{4}). Both the ${\cal E}(X)$ and the $\nabla_\al$ are anti-hermitean w.r.t. the scalar product Eqn.(\ref{4}).

The decomposability of ${\cal E}(X)$ w.r.t. to the operators $\nabla_\al$ will be crucial for the further development of the theory, especially for identifying the gauge field variables of the theory. 

Introducing the variation $\de_{_{\cal E}} ..\equiv ..' - ..$ of an expression under a gauge transformation we can write
\beq \label{11}
\de_{_{\cal E}}\psi (x,X) \equiv \psi' (x,X) - \psi (x,X) = -\, {\cal E}^\al (X) \cdot \nabla_\al\,\psi (x,X). \eeq

The variation of the Lagrangian density ${\cal L}_M(\psi, \pa_\mu \psi)$ - depending on $x$ and $X$ only through the fields $\psi$ and their $x$-derivatives $\pa_\mu \psi$ - becomes
\beq \label{12}
\de_{_{\cal E}} {\cal L}_M (\psi, \pa_\mu \psi) = -\, {\cal E}^\al \cdot \nabla_\al\, {\cal L}_M (\psi, \pa_\mu \psi)
\eeq
implying the global invariance of the corresponding Lagrangian
\beqq \label{13}
\de_{_{\cal E}} L_M &=& \intX \La^{-4} \de_{_{\cal E}} {\cal L}_M (\psi, \pa_\mu \psi) \nonumber \\
&=& - \intX \La^{-4}
\,\nabla_\al \left({\cal E}^\al {\cal L}_M (\psi, \pa_\mu \psi)\right) = 0. 
\eeqq
Here we have used the unimodularity condition $\nabla_\al {\cal E}^\al = 0$ so that the $\intX$-integration yields zero for fields $\psi$ and gauge parameters ${\cal E}$ vanishing at infinity in $X$-space. As a result any matter Lagrangian is {\it automatically} globally gauge invariant under volume-preserving diffeomorphisms.

The invariance Eqn.(\ref{13}) implies the existence of four conserved Noether currents
\beqq \label{14} J^\nu\,_\al &\equiv& \intX \La^{-4}\, \frac{\pa{\cal L}_M}{\pa(\pa_\nu \psi)} \nabla_\al\, \psi \nonumber \\
\pa_\nu J^\nu\,_\al &=& 0,\:\:\al=\sl{0,1,2,3}  \eeqq
and the four time-independent charges
\beq \label{15} {\bf P}_\al\equiv \int\! d^{\sl 3}x \, J^{\sl 0}\,_\al, \:\:\al=\sl{0,1,2,3} \eeq
which generate the inner global coordinate transformations in field space. It is these currents and charges which will be interpreted in terms of gravitational energy-momentum and become the sources of ${\overline{\bf diff}}\,{\bf R}^{\sl 4}$ gauge fields.

\section{Local Diffeomorphism Invariance, Covariant Derivatives and Gauge Fields}
In this section we introduce local gauge transformations and - to make globally invariant Lagrangians locally invariant - the corresponding covariant derivatives, gauge field and covariant field strength operators. We also define global inner scale transformations under which the covariant derivative, gauge field and covariant field strength operators are invariant.

Let us extend the global volume-preserving diffeomorphism group represented in field space to a group of local transformations by allowing ${\cal E}^\al (X)$ to vary with $x$ as well, i.e. allowing for $x$-dependent volume-preserving general coordinate transformations ${\cal E}^\al (X)\ar {\cal E}^\al (x,X)$ in inner space. In other words the group we gauge is the group of all isometric diffeomorphisms preserving the volume in inner space.

In generalization of Eqn.(\ref{10}) we thus consider
\beq \label{16}
U_{\cal E} (x,X) \equiv {\bf 1} - {\cal E}(x,X),\:\: {\cal E}(x,X)= {\cal E}^\al (x,X)\cdot \nabla_\al.
\eeq
The formulae Eqns.(\ref{6}) together with Eqn.(\ref{7}) still define the representation of the volume-preserving diffeomorphism group in field space.

To assure local gauge covariance for globally diffeomorphism covariant Lagrangian densities as in Eqn.(\ref{12}) we must introduce a covariant derivative $D_\mu$ which is defined by the transformation requirement
\beq \label{17}
D'_\mu \,\, U_{\cal E} (x,X) = U_{\cal E} (x,X)\,\, D_\mu,
\eeq
where $D'_\mu$ denotes the gauge-transformed covariant derivative.

By construction the Lagrangian density in Eqn.(\ref{5}) with covariant derivatives replacing the ordinary ones $\pa_\mu\ar D_\mu$ transforms covariantly under local infinitesimal transformations
\beq \label{18}
\de_{_{\cal E}} {\cal L}_M (\psi, D_\mu \psi)  = - {\cal E}^\al (x,X)\cdot \nabla_\al \, {\cal L}_M(\psi, D_\mu \psi)
\eeq
and the corresponding Lagrangian is locally gauge invariant
\beq \label{19}
\de_{_{\cal E}} L_M(\psi, D_\mu \psi)
= - \intX \La^{-4} \nabla_\al \left({\cal E}^\al (x,X) {\cal L}_M(\psi, D_\mu \psi) \right)
= 0
\eeq
again due to the unimodularity condition $\nabla_\be {\cal E}^\be (x,X)=0$.

Next, to fulfil Eqn.(\ref{17}) we make the usual ansatz
\beq \label{20}
D_\mu (x,X) \equiv \pa_\mu + A_\mu (x,X),\quad A_\mu (x,X) \equiv A_\mu\,^\al (x,X)\cdot \nabla_\al
\eeq
decomposing $A_\mu (x,X)$ w.r.t the generators $\nabla_\al$ of the diffeomorphism algebra in field space. In order to have the gauge fields in the algebra ${\overline{\bf diff}}\,{\bf R}^{\sl 4}$ we impose in addition
\beq \label{21}
\nabla_\al A_\mu\,^\al (x,X) = 0
\eeq
consistent with $\nabla_\be {\cal E}^\be (x,X)=0$. As a consequence the usual ordering problem for $A_\mu\,^\al$ and $\nabla_\al$ in the definition of $D_\mu$ does not arise and $D_\mu$ is anti-hermitean w.r.t to the scalar product defined above.

The requirement Eqn.(\ref{17}) translates into the transformation law for the gauge field
\beq \label{22}
\de_{_{\cal E}} A_\mu (x,X) = \pa_\mu {\cal E} (x,X) -  \left[{\cal E} (x,X), A_\mu (x,X) \right] \eeq
which reads in components
\beq \label{23}
\de_{_{\cal E}} A_\mu\,^\al = \pa_\mu {\cal E}^\al + A_\mu\,^\be \cdot \nabla_\be {\cal E}^\al - {\cal E}^\be \cdot \nabla_\be A_\mu\,^\al 
\eeq
respecting $\nabla_\al \de_{_{\cal E}} A_\mu\,^\al = 0$. The inhomogenous term $\pa_\mu {\cal E}^\al$ assures the desired transformation behaviour of the $D_\mu$, the term $\nabla_\be {\cal E}^\al \cdot A_\mu\,^\be $ rotates the inner space vector $A_\mu\,^\be$ and the term $-\, {\cal E}^\be \cdot \nabla_\be A_\mu\,^\al$ shifts the coordinates $X^\al \ar X'^\al = X^\al + {\cal E}^\al (x,X)$.

Note that the consistent decomposition of both $A_\mu$ and $A'_\mu$ w.r.t. the generators $\nabla_\al$ is crucial for the theory's viability. It is ensured by the closure of the algebra Eqn.(\ref{8}) and the gauge invariance of $\nabla_\al A_\mu\,^\al =0$ for gauge parameters fulfilling $\nabla_\be {\cal E}^\be =0$.

Let us next define the field strength operator $F_{\mu\nu}$ in the usual way
\beqq \label{24}
F_{\mu\nu}(x,X)&\equiv& \left[D_\mu (x,X), D_\nu (x,X) \right] \nonumber \\
&=& F_{\mu\nu}\,^\al (x,X)\cdot \nabla_\al 
\eeqq
which again can be decomposed consistently w.r.t. $\nabla_\al$.
The field strength components $ F_{\mu\nu}\,^\al (x,X)$ are calculated to be
\beqq \label{25}
F_{\mu\nu}\,^\al (x,X)
&\equiv& \pa_\mu A_\nu\,^\al (x,X) - \pa_\nu A_\mu\,^\al (x,X) \nonumber \\
&+& A_\mu\,^\be (x,X)\cdot \nabla_\be A_\nu\,^\al (x,X) \\
&-& A_\nu\,^\be (x,X)\cdot \nabla_\be A_\mu\,^\al (x,X). \nonumber
\eeqq
Under a local gauge transformation the field strength and its components transform covariantly
\beqq \label{26}
\de_{_{\cal E}} F_{\mu\nu} (x,X) &=& - \left[{\cal E} (x,X) , F_{\mu\nu} (x,X) \right], \nonumber \\
\de_{_{\cal E}} F_{\mu\nu}\,^\al &=& F_{\mu\nu}\,^\be \cdot \nabla_\be {\cal E}^\al - {\cal E}^\be \cdot \nabla_\be F_{\mu\nu}\,^\al. 
\eeqq
As required for algebra elements $\nabla_\al F_{\mu\nu}\,^\al = 0$ and $\nabla_\al \de_{_{\cal E}} F_{\mu\nu}\,^\al = 0$ for gauge fields fulfilling $\nabla_\al A_\mu\,^\al = 0$ and gauge parameters fulfilling $\nabla_\be {\cal E}^\be =0$.

Besides the global and local invariance under inner coordinate transformations Eqns.(\ref{6}) the theory features another global invariance in inner space - namely scale invariance.
Let us give the respective transformation law for a rescaling with scale parameter $\rho\in {\bf R}^+$
\beqq \label{27}
x^\mu \lar x'^\mu = x^\mu, \quad 
X^\al &\lar& X'^\al = \rho X^\al, \quad
\La \lar \La' = \rho \La,  \nonumber \\
\psi (x,X) &\lar& \psi' (x,X') = \psi (x,X), \\
A_\mu\,^\al (x,X) &\lar& A'_\mu\,^\al (x,X') = \rho A_\mu\,^\al (x,X), \nonumber \\
F_{\mu\nu}\,^\al (x,X) &\lar& F'_{\mu\nu}\,^\al (x,X') = \rho F_{\mu\nu}\,^\al (x,X).
\nonumber
\eeqq
Under Eqns.(\ref{27}) matter Lagrangians and the operators $D_\mu$, $A_\mu$ and $F_{\mu\nu}$ are invariant which will prove crucial to consistently define the theory below.

\section{The Lagrangian}
In this section we introduce a metric in the inner space and derive the gauge field Lagrangian minimal in the sense of being gauge-invariant and of lowest possible dimension in the fields.

As heuristically motivated by analogy to the Yang-Mills case we propose the local gauge field Lagrangian to be proportional to $\Tr F^2$ - ensuring gauge invariance and at most second order dependence on the first derivatives of the $A_\mu\,^\al$-fields which is crucial for a quantization leading to a unitary and renormalizable theory.

To make sense of the formal operation $\Tr$ and to define $\Tr F^2$ properly let us start with the evaluation of the differential operator product
\beq \label{28}
F_{\mu\nu}\, F^{\mu\nu} = F_{\mu\nu}\,^\al F^{\mu\nu\,\be} {\nabla\rvec}_\al {\nabla\rvec}_\be + 
\nabla_\al \left(F_{\mu\nu}\,^\al F^{\mu\nu\,\be}\right) {\nabla\rvec}_\be,
\eeq
where ${\nabla\rvec}_\be$ acts on all fields to its right.

To be able to evaluate the trace in a coordinate system we would like to insert complete systems of $X$- and $P$-vectors
\beq \label{29}
{\bf 1} = \intX \mid \!X\rangle\langle X\!\mid\, , \quad\quad
{\bf 1} = \intP \mid \!P\rangle\langle P\!\mid
\eeq
under the $\Tr$-operation and using $\langle X\!\mid\! P\rangle=\exp(i\, P\cdot X)$. This assumes, however, the existence of Cartesian coordinates and a metric in inner space and the existence of both co- and contravariant vectors w.r.t. that metric.

So let us endow the inner four-dimensional real vector space ${\bf R}^{\sl 4}$ with a metric $g_{\al\be}(x,X)$ of Minkowskian signature and require that its geometry - which we take as an a priori - is flat, Riem$(g) = 0$. This means that it is always possible to choose global Cartesian coordinates with the metric $g_{\al\be}(x,X) = \eta_{\al\be}$ collapsing to the global Minkowski metric. Such choices of coordinates amount to partially fixing a gauge and we will call them {\it Minkowskian gauges} in the following.

Note that under inner coordinate transformations the metric transforms as a contravariant tensor
\beq \label{30}
\de_{_{\cal E}} g_{\al\be}= - {\cal E}^\ga \cdot \nabla_\ga g_{\al\be} - g_{\ga\be}\cdot \nabla_\al {\cal E}^\ga - g_{\al\ga}\cdot \nabla_\be {\cal E}^\ga.
\eeq 

Working in Cartesian coordinates we can now insert complete systems of $X$- and $P$-vectors and formally take the trace over the inner space
\beqq \label{31}
\Tr \Big\{F_{\mu\nu}\, F^{\mu\nu}\Big\}_\eta
&\propto& \intX\intP \langle X\!\mid F_{\mu\nu}\, F^{\mu\nu}
\mid \!P\rangle\langle P\!\mid\! X\rangle \nonumber \\
&=& \intX\intP \Bigg\{ F_{\mu\nu}\,^\al\cdot
F^{\mu\nu\,\be} \,i P_\al \,i P_\be \\
&+& \nabla_\al \left(F_{\mu\nu}\,^\al F^{\mu\nu\,\be}\right) i P_\be \Bigg\} \nonumber \eeqq
which has still to be properly defined. Above we have made use of $\nabla_\al\!\mid \!P\rangle = i P_\al \!\mid \!P\rangle$ and the subscript $\Tr \{\dots\}_\eta$ denotes evaluation in a given coordinate system and for a given metric, in this case Cartesian coordinates and the Minkowski metric. Note that beeing a total divergence in $X$-space and odd in $P$ the second term in Eqn.(\ref{31}) vanishes.

The definition of the remaining $P$-integral requires care in order to covariantly deal with the infinities resulting from the non-compactness of the gauge group. Noting that the regularization will restrict $P$ to the forward and backward light cones, i.e. $- P^2 \geq 0$ we extract the tensor structure
\beq \label{32} 
\intP\, (- P_\al P_\be) = \frac{1}{4}\, \La^{-6}\, \Om_{\sl 1}\, \eta_{\al\be},
\eeq
and isolate the infinity into a dimensionless Lorentz-invariant integral of the type
\beq \label{33}
\Om_n \sim \ \La^{2n + 4} \intP\, (-P^2)^n,
\eeq
where the subscript counts powers of $-P^2$. Slicing the inner Minkowski space into light-like, time-like and space-like shells of invariant lengths $-P^2 = M^2,\:\: -\infty \leq M^2 \leq \infty$ which are invariant under proper Lorentz transformations we can identically rewrite
\beq \label{34}
\Om_n \sim \La^{2n + 4} \int_{-\infty}^{\infty} dM^2\, M^{2n} \intP\, \de\!\left(M^2 + P^2\right).
\eeq
To regularize $\Om_n$ in a Lorentz-invariant way we first cut off the space-like shells with negative lengths $M^2 < 0$ and split $1 = \theta(P^{\sl 0}) + \theta(- P^{\sl 0})$ so that
\beqq \label{35}
\Om_n &\sim& \La^{2n + 4} \int_0^{\infty} dM^2\, M^{2n}
\intP\, \de\!\left(M^2 + P^2\right) \\
& & \quad\quad\quad \cdot\left(\theta(P^{\sl 0}) + \theta(- P^{\sl 0})\right) \nonumber
\eeqq
which is a Lorentz-invariant procedure. As we will see in the section on Hamiltonian field dynamics this cutoff arises naturally from the condition of positivity for the Hamiltonian which will restrict all fields Fourier-transformed over inner space to have support on the set ${\bf V^+}(P)\cup {\bf V^-}(P)$, where
\beq \label{36}
{\bf V^\pm}(P) = \{P\in {\bf M^{\sl 4}}\mid -P^2 \geq 0,\: \pm P^{\sl 0} \geq 0\}
\eeq
denote the forward and backward light cones.

Second, there is always a Lorentz frame with a time-like vector $L^\al$ which has $L^2 = -\La^{-2}$ as its invariant length so that $L^\al = (\La^{-1},\underline 0)$ in this frame.
Third, 
\beq \label{37}
-L^2 \pm 2\, L\cdot P = \La^{-2} \mp 2\,\La^{-1}\,P^{\sl 0}
\eeq
are Lorentz scalars.

This allows us to {\it define} $\Om^\La_n$ as an integral over of the forward cone ${\bf V^+}(P)$ with a cutoff for $P^{\sl 0} = \sqrt{M^2 + {\underline P}^2} \leq \frac{1}{2\,\La}$ and the backward cone ${\bf V^-}(P)$ with a cutoff for $P^{\sl 0} = -\sqrt{M^2 + {\underline P}^2} \geq -\frac{1}{2\,\La}$ for fixed $M$ first and then summing over all $M \leq \frac{1}{2\,\La}$
\beqq \label{38}
\Om^\La_n &\equiv& \La^{2n + 4} \int_0^{\frac{1}{4\, \La^2}}
\!\!\!\! dM^2\, M^{2n} \intP\, \de\!\left(M^2 + P^2 \right)
\nonumber \\
& & \cdot\left(
\theta(P^{\sl 0}) \theta(-L^2 + 2\, L\cdot P) +
\theta(- P^{\sl 0}) \theta(-L^2 - 2\, L\cdot P)\right) \\
&=& \frac{1}{(2\pi)^3\,4^{n + 2}}\, \int_0^1
\!\! dx\, x^n\, \left(\sqrt{1 - x} - x \ln
\frac{\sqrt{1 - x} + 1}{\sqrt x} \right)
\nonumber
\eeqq
which is a positive and finite Lorentz scalar for all $n$. Explicitly we find $\Om^\La_{\sl 1} = \frac{1}{720\, (4\pi)^3}$. As $\La^{-1}$ is the only a priori mass scale in the theory any other Lorentz-invariant regularization procedure just changes the numerical values of $\Om_n$.

Note that regularized in this way any inner $P$-integral over polynomials in $P$ reduces to products of the metric in inner space and $\Om^\La_n$s and is as well behaved as the usual sums over structure constants of a compact Lie group are in a Yang-Mills theory.

Using Eqns.(\ref{32}) and (\ref{38}) to evaluate Eqn.(\ref{31}) we now {\it define} a $\La$-dependent trace in Minkowskian gauges by
\beq \label{39}
\Tr_{\La} \Big\{F_{\mu\nu}\, F^{\mu\nu}\Big\}_\eta
= \frac{\Om^\La_{\sl 1}}{4} \intX \La^{-4} \, \La^{-1} F_{\mu\nu}\,^\al
\cdot \La^{-1} F^{\mu\nu}\,_\al 
\eeq
which is easily generalized to arbitrary coordinates in inner space
\beq \label{40}
\Tr_\La \Big\{F_{\mu\nu}\, F^{\mu\nu}\Big\}_g
= \frac{\Om^\La_{\sl 1}}{4} \intX \La^{-4} \, \La^{-1} F_{\mu\nu}\,^\al
\cdot \La^{-1} F^{\mu\nu}\,_\al, 
\eeq
where we have to contract the inner indices with $g$ now. The expression above is obviously well defined in any coordinate system and gauge-invariant under the combined transformations of field strenght components Eqns.(\ref{26}) and the metric Eqn.(\ref{30}).

Finally this allows us to write down the Lagrangian for the gauge fields 
\beq \label{41}
L (A_\nu\,^\al, \pa_\mu A_\nu\,^\al, \nabla_N A_\nu\,^\al, \La)
\equiv -\frac{1}{\Om^\La_{\sl 1}}\,\Tr_\La \Big\{F_{\mu\nu}\, F^{\mu\nu}\Big\}_g
\eeq
and the corresponding Lagrangian density
\beq \label{42}
{\cal L} (A_\nu\,^\al, \pa_\mu A_\nu\,^\al, \nabla_N A_\nu\,^\al, \La)
= -\frac{1}{4\, \La^2} \, F_{\mu\nu}\,^\al \cdot F^{\mu\nu}\,_\al.
\eeq
Both are dimensionless in inner space - the Lagrangian density due to the factors of $\La$. Note that the factor of $\frac{1}{4} $ above leads to the usual normalization of the quadratic part of the Lagrangian density and the overall minus sign will yield a positive Hamiltonian as we will show in the section on Hamiltonian field dynamics.

Note that the Lagrangian for $\rho\La$ is related to the Lagrangian for a given $\La$ by
\beq \label{43}
L (\rho X,\rho A_\nu\,^\al (X),\dots,\rho \La) =
L (X,A_\nu\,^\al (X),\dots, \La) 
\eeq
with a similar relation holding for the matter Lagrangian Eqn.(\ref{5}) - the dependence of the theory on $\La$ is controlled by the scale transformation Eqn.(\ref{27}). In other words {\it theories for different $\La$ are equivalent up to inner rescalings}.

This is a crucial point which will allow us to rescale $\La$ always to the Planck length, a fact we will use when extracting the physics of the theory at hands.

Why have we not simply written down Eqn.(\ref{42})? First, the calculation starting with the $\Tr$-operation shows that the dimensionful parameter $\La$ automatically emerges in the definition of the Lagrangian and that the theory at $\La$ is related in a simple way to the one at $\rho \La$. We would not have uncovered this somewhat hidden, but crucial fact in simply writing down the Lagrangian. Second, we will have to show in the quantized version of the theory that the kinematic integrals generalizing the kinematic sums over gauge degrees of freedom in the Yang-Mills case can be consistently defined. The definition of $\Tr F^2$ is a first example of how this will be achieved.

\section{Lagrangian Field Dynamics}
In this section we develop the Lagrangian field dynamics determining the field equations which will not depend on the metric $g$ and derive the most important conservation laws for the theory. 

Note that by definition we always work with fields living in the algebra ${\overline{\bf diff}}\,{\bf M }^{\sl 4}$ from now on. We start with the action
\beq \label{44}
S = -\frac{1}{\Om^\La_{\sl 1}}\,\intx \,\Tr_\La \Big\{F_{\mu\nu}\, F^{\mu\nu}\Big\}_g.
\eeq
Variation of Eqn.(\ref{44}) w.r.t. $A^{\nu\,\be}$ to get a stationary point 
\beqq \label{45}
\de S &=& -\frac{4}{\Om^\La_{\sl 1} }\,\intx \,\Tr_\La \Bigg\{ \Big(
-\pa^\mu F_{\mu\nu}\,^\al {\nabla\rvec}_\al \nonumber \\
&+& F_{\mu\nu}\,^\ga {\nabla\rvec}_\ga A^{\mu\,\al} {\nabla\rvec}_\al
- A^{\mu\,\ga} {\nabla\rvec}_\ga F_{\mu\nu}\,^\al {\nabla\rvec}_\al \Big)
\, \de A^{\nu\,\be} {\nabla\rvec}_\be 
\Bigg\}_g \\
&=& -\frac{4}{\Om^\La_{\sl 1}}\,\intx \,\Tr_\La \Bigg\{\Big(
-\pa^\mu F_{\mu\nu}\,^\al \nonumber \\
&-& A^{\mu\,\ga}\cdot \nabla_\ga F_{\mu\nu}\,^\al 
+ F_{\mu\nu}\,^\ga\cdot \nabla_\ga A^{\mu\,\al} \Big) \, \de A^{\nu\,\be} 
{\nabla\rvec}_\be {\nabla\rvec}_\al \Bigg\}_g \nonumber \\
&=& 0 \nonumber
\eeqq
yields the field equations
\beq \label{46}
\pa^\mu F_{\mu\nu}\,^\al + A^{\mu\,\be}\cdot \nabla_\be F_{\mu\nu}\,^\al 
- F_{\mu\nu}\,^\be\cdot \nabla_\be A^{\mu\,\al} = 0
\eeq
which by inspection do not depend on the metric. This means that the metric $g$ is not an independent dynamical field and irrelevant for the dynamics of the gauge fields. Above we have used the cyclicality of the trace, partially integrated and brought all the ${\nabla\rvec}_\al$ to the right. Note that under the trace all terms with an odd number of ${\nabla\rvec}_\al$ vanish.

The equations of motion can be brought into a covariant form
\beq \label{47}
{\cal D}^{\mu\, \al}\,_\be F_{\mu\nu}\,^\be = 0
\eeq
introducing the covariant derivative ${\cal D}^{\mu\, \al}\,_\be$ acting on vectors in inner space
\beq \label{48}
{\cal D}^{\mu\, \al}\,_\be \equiv \pa^\mu\, \de^\al\,_\be + A^{\mu\, \ga} \cdot \nabla_\ga\, \de^\al\,_\be - \nabla_\be A^{\mu\,^\al}.
\eeq 

By inspection the covariant derivative Eqn.(\ref{48}) respects the gauge algebra and is an endomorphism of ${\overline{\bf diff}}\,{\bf M}^{\sl 4}$ because
\beq \label{49}
\nabla_\al {\cal D}^{\mu\, \al}\,_\be G^\be = 0
\eeq
for $\nabla_\be G^\be = 0$.

Finally we can recast the field equations in coordinate-independent and manifestly covariant form
\beq \label{50}
\left[D_\mu, F^{\mu\nu}\right] = 0
\eeq
underlining the formal similarity of the present theory to Yang-Mills theories of compact Lie groups.

The $4\times 4$ field equations Eqns.(\ref{46}) clearly display the self coupling of the $A^\mu\,_\al$-fields to the four conserved Noether current densities
\beqq \label{51}
& & {\cal J}_\nu\,^\al \equiv
A^{\mu\,\be}\cdot \nabla_\be F_{\mu\nu}\,^\al
- F_{\mu\nu}\,^\be\cdot \nabla_\be A^{\mu\,\al} \nonumber \\
& & \quad\quad \pa^\nu {\cal J}_\nu\,^\al = 0,\:\: \al=\sl{0,1,2,3}  \eeqq
which obey the restrictions on algebra elements $\nabla_\al {\cal J}_\nu\,^\al = 0$ as expected.

Next we analyze the invariance of the action Eqn.(\ref{44}) under spacetime translations and derive the conserved energy-momentum tensor. In the usual way we obtain the canonical energy-momentum tensor
\beq \label{52}
T^\mu\,_\nu = \frac{4}{\Om^\La_{\sl 1}}\, \Tr_\La \left\{\frac{1}{4}\,\eta^\mu\,_\nu\,F_{\rho\si}\, F^{\rho\si}
- F^{\mu\rho}\, \pa_\nu A_\rho \right\}_g
\eeq
which is conserved $\pa_\mu T^\mu\,_\nu = 0$. As in other gauge field theories this tensor is, however, not gauge invariant. Using the field equations Eqns.(\ref{50}) and the cyclicality of the trace we find
\beq \label{53}
\pa_\rho \Tr_\La \left\{F^{\mu\rho}\, A_\nu \right\}_g = \Tr_\La \left\{F^{\mu\rho}\,\left(\pa_\rho A_\nu + [A_\rho,A_\nu]\right) \right\}_g.
\eeq
Adding this total divergence we finally get an improved, conserved and gauge-invariant energy-momentum tensor
\beqq \label{54} 
\Th^\mu\,_\nu
&=& T^\mu\,_\nu +\,\frac{4}{\Om^\La_{\sl 1}}\, \pa_\rho \Tr_\La \left\{F^{\mu\rho}\, A_\nu \right\}_g \nonumber \\
&=& \frac{4}{\Om^\La_{\sl 1}}\, \Tr_\La \left\{\frac{1}{4}\,\eta^\mu\,_\nu\,F_{\rho\si}\, F^{\rho\si}
- F^{\mu\rho}\, F_{\nu\rho}\right\}_g 
\eeqq
which reads in components
\beq \label{55}
\Th^\mu\,_\nu = \intX \La^{-6} \left\{\frac{1}{4}\,\eta^\mu\,_\nu\, F_{\rho\si}\,^\al \cdot F^{\rho\si}\,_\al - \,F^{\mu\rho}\,_\al \cdot F_{\nu \rho}\,^\al \right\}.
\eeq
The corresponding time-independent momentum four-vector reads
\beq \label{56}
{\bf p}_\mu\equiv \int\! d^{\sl 3}x \Th^{\sl 0}\,_\mu
\eeq
and generates the translations in spacetime.

In addition, the theory is obviously Lorentz and - at the classical level - scale invariant under the corresponding spacetime and field transformations. We do not display the corresponding conserved currents and charges here.

Let us finally write down the Bianchi identities
\beq \label{57} {\cal D}_\rho^\al\,_\be F_{\mu\nu}\,^\be + \;\mbox{cyclical in}\; \rho,\mu,\nu = 0. \eeq

The equations above define a perfectly consistent classical dynamical system within the Lagrangian framework. Note that for physical observables such as the energy-momentum tensor the inner degrees of freedom are integrated over.

As we ultimately aim at quantizing the theory we next turn to develop the Hamiltonian field theory.

\section{Hamiltonian Field Dynamics}
In this section we develop the Hamiltonian field dynamics closely following \cite{stw2}. We fix a gauge first choosing Cartesian coordinates along with the Minkowski metric in inner space and second eliminating the first class constraints related to the remaining gauge degrees of freedom by imposing the axial gauge condition. We then explicitely solve the remaining constraints and find the unconstrained canonical variables for the theory. Finally we reexpress the gauge-fixed Hamiltonian $H$ of the theory in these variables displaying its positivity explicitly. This will serve in \cite{chw3} as the starting point for quantization.

Let us start using the gauge freedom of the theory to choose Cartesian coordinates along with the Minkowski metric in inner space, i.e. fixing a gauge up to coordinate transformations Eqns.(\ref{16}) which leave the Minkowski metric invariant. The remaining gauge group is just the Poincar\'e group acting on the inner Minkowski space with infinitesmal parameters ${\cal E}^\al (x,X) = \ep^\al (x) +
\om^\al\,_\be (x)\, X^\be,\:\: \om_{\al\be} = -\om_{\be\al }$.

Hence, we start with the Lagrangian density Eqn.(\ref{42})
\beq \label{58}
{\cal L} (A_\nu\,^\al, \pa_\mu A_\nu\,^\al, \nabla_\be A_\nu\,^\al, \La)
= -\frac{1}{4\, \La^2} \, F_{\mu\nu}\,^\al \cdot F^{\mu\nu}\,_\al,
\eeq
where the $A_\nu\,^\al$ are the gauge fields, $F_{\mu\nu}\,^\al$ are given by Eqn.(\ref{25}) and where the $\al$-indices are raised and lowered with $\eta_{\al\be}$.

Next we define the variables $\Pit^\mu\,\!_\al$ conjugate to $A_\mu\,^\al$ by
\beq \label{59}
\Pit^\mu\,\!_\al
\equiv \La\, \frac{\pa{\cal L}}{\pa(\pa_{\sl 0} A_\mu\,^\al)}
= -\frac{1}{\La}\, F^{{\sl 0}\mu}\,_\al
\eeq
which are dimensionless in inner space. By definition they are elements of the gauge algebra ${\overline{\bf diff}}\,{\bf M }^{\sl 4}$ and fulfil
\beq \label{60}
\nabla^\al \Pit^j\,\!_\al = 0,\:\: j=\sl{1,2,3}.
\eeq

As usual we find the two sets of four constraints
\beq \label{61}
\Pit^{\sl 0}\,\!_\al = 0
\eeq
and
\beq \label{62}
\pa_k\, \Pit^k\,\!_\al + A_k\,^\be\cdot \nabla_\be\, \Pit^k\,\!_\al - \Pit^k\,\!_\be\cdot \nabla^\be A_{k\,\al} = 0
\eeq
which are the field equations Eqn.(\ref{46}) for $\nu = {\sl 0}$ and $k=\sl{1,2,3}$. 

The Poisson brackets of the two constraints Eqns.(\ref{61}) and (\ref{62}) w.r.t  $A_i\,^\al$, $\Pit^j\,_\be$ vanish because Eqn.(\ref{62}) is independent of $A_{\sl 0}\,^\al$. Hence, they are first class. To properly deal with them we fully fix the remaining gauge degrees of freedom - Poincar\'e transformations which leave the Minkowski metric invariant - by imposing the axial gauge condition
\beq \label{63}
A_{\sl 3}\,\!^\al =0.
\eeq

The canonical variables of the theory reduce to $A_i\,^\al$ and their conjugates $\Pit_j\,\!^\be$ 
\beqq \label{64}
\Pit_j\,\!^\be &=& \frac{1}{\La}\, F_{{\sl 0}j}\,^\be 
= \frac{1}{\La}\,\Big(\pa_{\sl 0} A_j\,^\be - \pa_j A_{\sl 0}\,^\be \nonumber \\
&+& A_{\sl 0}\,^\al \cdot \nabla_\al A_j\,^\be - A_j\,^\al \cdot \nabla_\al A_{\sl 0}\,^\be \Big)
\eeqq
for $i,j = {\sl 1,2}$ only.

$A_{\sl 0}\,^\al$ is not an independent variable, but can be expressed in terms of the canonical variables above by solving the constraint Eqn.(\ref{62})
\beqq  \label{65}
A_{\sl 0}\,^\al &=& \frac{1}{\pa_{\sl 3}\,\!^2} \, \La\, \sum_{i={\sl 1}}^{\sl 2} \left(\pa_i \Pit_i\,\!^\al + A_i\,^\be\cdot \nabla_\be \Pit_i\,\!^\al - \Pit_i\,\!^\be \cdot \nabla_\be A_i\,^\al \right) \nonumber \\
&=& \frac{1}{\pa_{\sl 3}\,\!^2}\, \La\, \sum_{i={\sl 1}}^{\sl 2} {\cal D}_i^\al\,_\be \Pit_i\,\!^\be,
\eeqq
where we have used $F_{{\sl 30}}\,^\al = \pa_{\sl 3} A_{\sl 0}\,^\al$.

Finally we solve the unimodularity constraints on $A_i\,^\al$
\beq  \label{66}
\nabla_\al A_i\,^\al = \nabla_{\sl 0} A_i\,^{\sl 0} + \nabla_a A_i\,^a
= 0,\:\: i = {\sl 1,2}, 
\eeq
where $a$ runs over $a = {\sl 1,2,3}$ to obtain
\beqq  \label{67}
A_i\,^{\sl 0} (x,X) &=& -\int^{X^{\sl 0}}\!\!\! dS\, \nabla_a A_i\,^a (x;S\, ,X^{\sl 1},X^{\sl 2},X^{\sl 3}) \nonumber \\
&\equiv& -\frac{1}{\nabla_{\sl 0}}\, \nabla_a A_i\,^a,
\:\: i = {\sl 1,2} 
\eeqq
and analogously for $\Pit_j\,\!^\be$
\beq  \label{68}
\Pit_j\,\!^{\sl 0} (x,X) = -\frac{1}{\nabla_{\sl 0}}\,
\nabla_a \Pit_j\,\!^a ,\:\: i = {\sl 1,2} 
\eeq
further reducing the independent variables to
$A_{\sl 1}\,^{\sl 1}$,
$A_{\sl 1}\,^{\sl 2}$,
$A_{\sl 1}\,^{\sl 3}$,
$A_{\sl 2}\,^{\sl 1}$,
$A_{\sl 2}\,^{\sl 2}$,
$A_{\sl 2}\,^{\sl 3}$
and the respective $\Pit_j\,\!^a$ s.

The Hamiltonian in the original gauge field variables
\beq \label{69}
H \equiv \int\! d^{\sl 3}x \intX \La^{-4}\, \La^{-1}\,  \Pit^\mu\,_\al\cdot \pa_{\sl 0}A_\mu\,^\al - L
\eeq
reduces in the axial gauge to
\beqq \label{70}
H
&=& \int\! d^{\sl 3}x \intX \La^{-4}\, \La^{-1}\, \sum_{i={\sl 1}}^{\sl 2} \Pit_{i\al}\cdot \pa_{\sl 0}A_i\,^\al - L \nonumber \\
&=& \int\! d^{\sl 3}x \intX \La^{-4}\, \Bigg\{
\frac{1}{2\,\La^2}\, \pa_{\sl 3}A_{\sl 0}\,^\al \cdot \pa_{\sl 3}A_{{\sl 0}\al} \\
&+& \frac{1}{2}\, \sum_{i={\sl 1}}^{\sl 2} \Pit_i\,\!^\al \cdot \Pit_{i\al}  
+ \frac{1}{4\,\La^2}\, \sum_{i,j={\sl 1}}^{\sl 2} F_{ij}\,^\al \cdot F_{ij\,\al}
\nonumber \\
&+& \frac{1}{2\,\La^2}\, \sum_{i={\sl 1}}^{\sl 2} \pa_{\sl 3}A_i\,^\al \cdot \pa_{\sl 3}A_{i\al} \Bigg\}, \nonumber
\eeqq
where we have made use of Eqns.(\ref{62}) and (\ref{64}) to rearrange terms and where $A_{\sl 0}\,^\al$ is given by Eqn.(\ref{65}). Note that $H = {\bf p}_{\sl 0}$ as expected.

The use of Eqn.(\ref{67}) allows us next to rewrite
\beqq \label{71}
\intX \La^{-4}\, \pa_{\sl 3}A_i\,^\al \cdot \pa_{\sl 3}A_{i\,\al}
&=& \intX \La^{-4}\, \Bigg\{\pa_{\sl 3}A_i\,^a\cdot \pa_{\sl 3}A_{i\,a} \nonumber \\
&-&\pa_{\sl 3}\, \frac{1}{\nabla_{\sl 0}}\, \nabla_a A_i\,^a\cdot \pa_{\sl 3}\, \frac{1}{\nabla_{\sl 0}}\, \nabla_b A_i\,^b \Bigg\} \\
&=& \int\! d^{\sl 4}K\, \La^4\, \pa_{\sl 3}{\hat A^*}_i\,^a\cdot M_{ab} (K) \, \pa_{\sl 3}{\hat A}_i\,^b, \nonumber
\eeqq
where we have Fourier-transformed $A_i\,^a$ in inner space
\beq \label{72}
A_i\,^a (x,X) = \int\! \frac{d^{\sl 4}K}{(2{\pi})^2}\, \La^4\,
e^{i\, K\cdot X}\, {\hat A}_i\,^a (x,K), 
\eeq
used the reality condition on $A_i\,^a$
\beq \label{73}
A_i\,^a (x,X) = A^*_i\,^a (x,X) \Rightarrow 
{\hat A}_i\,^a (x,-K) = {\hat A^*}_i\,^a (x,K)
\eeq 
and introduced the matrix
\beq \label{74}
M_{ab} (K) \equiv \de_{ab}  - \frac{K_a\, K_b}{(K_{\sl 0})^2}.
\eeq 

$M(K)$ is real, symmetric with eigenvalues $1$, $1$ and $-\frac{K^2}{(K_{\sl 0})^2}$. Hence, there exists an orthogonal $3\times 3$-matrix $C (K)$, $C^T = C^{-1}$ which diagonalizes $ M (K)$:
$C\, M\, C^T = \mbox{diag} (1,1,-\frac{K^2}{(K_{\sl 0})^2})$. Rotating
\beq \label{75}
{\tilde{\hat A}}_i\,^a (x,K) =
C^a\,_b (K)\, {\hat A}_i\,^b (x,K)
\eeq
and using analogous expressions for all the terms appearing in Eqn.(\ref{70}) we finally get $H$ in terms of the unconstrained independent variables
${\tilde{\hat A}}_{\sl 1\,1}$,
${\tilde{\hat A}}_{\sl 1\,2}$,
${\tilde{\hat A}}_{\sl 1\,3}$,
${\tilde{\hat A}}_{\sl 2\,1}$,
${\tilde{\hat A}}_{\sl 2\,2}$,
${\tilde{\hat A}}_{\sl 2\,3}$ and the respective
$\Pitch_{j\,b}$ s
\beqq \label{76} 
H &=& \int\! d^{\sl 3}x \int\! d^{\sl 4}K\, \La^4\, \Bigg\{
\frac{1}{2\,\La^2}\, \sum_{a={\sl 1}}^{\sl 2} \pa_{\sl 3}{\tilde{\hat A^*}}_{{\sl 0}\,a} \cdot \pa_{\sl 3}{\tilde{\hat A}}_{{\sl 0}\,a} \nonumber \\
&+& \frac{1}{2}\, \sum_{i;a={\sl 1}}^{\sl 2} \Pitch_{i\,a} \cdot \Pitth_{i\,a}  
+ \frac{1}{4\,\La^2}\, \sum_{i,j;a={\sl 1}}^{\sl 2} {\tilde{\hat F^*}}_{ij\,a} \cdot {\tilde{\hat F}}_{ij\,a}
\nonumber \\
&+& \frac{1}{2\,\La^2}\, \sum_{i;a={\sl 1}}^{\sl 2}
\pa_{\sl 3}{\tilde{\hat A^*}}_{i\,a}\cdot \pa_{\sl 3}{\tilde{\hat A}}_{i\,a} \Bigg\} \\
& & +\int\! d^{\sl 3}x \int\! d^{\sl 4}K\, \La^4\, 
\left(\frac{-K^2}{(K_{\sl 0})^2}\right)\, \Bigg\{
\frac{1}{2\,\La^2}\, \pa_{\sl 3}{\tilde{\hat A^*}}_{{\sl 0\,3}} \cdot \pa_{\sl 3}{\tilde{\hat A}}_{{\sl 0\,3}} \nonumber \\
&+& \frac{1}{2}\, \sum_{i={\sl 1}}^{\sl 2} \Pitch_{i\,{\sl 3}} \cdot \Pitth_{i\,{\sl 3}}  
+ \frac{1}{4\,\La^2}\, \sum_{i,j={\sl 1}}^{\sl 2} {\tilde{\hat F^*}}_{ij\,{\sl 3}} \cdot {\tilde{\hat F}}_{ij\,{\sl 3}}
\nonumber \\
&+& \frac{1}{2\,\La^2}\, \sum_{i={\sl 1}}^{\sl 2}
\pa_{\sl 3}{\tilde{\hat A^*}}_{i\,{\sl 3}}\cdot \pa_{\sl 3}{\tilde{\hat A}}_{i\,{\sl 3}} \Bigg\}. \nonumber
\eeqq

We immediately recognize that positivity of $H$ is ensured by the independent field variables of the theory vanishing outside the set ${\bf V^+}(K)\cup {\bf V^-}(K)$, i.e.
\beq \label{77} 
\mbox{supp}\left({\tilde{\hat A}}_{i\,a}(K)\right)
= \mbox{supp}\left(\Pitch_{j\,b}(K)\right) 
\subseteq {\bf V^+}(K)\cup {\bf V^-}(K) \Rightarrow H\geq 0 
\eeq
which is obviously a Lorentz-invariant requirement. Hence we restrict all fields in inner $K$-space to have positive mass-squared support in ${\bf V^+}(K)\cup {\bf V^-}(K)$ which defines in turn the class of {\it admissible functional spaces} for the fields of the theory. Note that the Fourier-transformed fields constant in $X$-space have support $\{K = 0\} = {\bf V^+}(K)\cap {\bf V^-}(K)$ and hence positive $H$.

To specify the dynamics we finally write down the equal-time Poisson brackets for the unconstrained independent field variables ${\tilde{\hat A}}_{i\,a}$ and $\Pitch_{j\,b}$ with positive mass-squared support
\beq \label{78} 
\left\{
{\tilde{\hat A}}_{i\,a} (x,K), \Pitch_{j\,b} (y,Q) \right\}_{x^{\sl 0} = y^{\sl 0}}
= \de_{ij}\, \de_{ab}\, \La^{-4}\, \de^{\sl 4}(K - Q)\, \de^{\sl 3}({\underline x} - {\underline y}),
\eeq
where $i,j = {\sl 1,2}$ and $a,b = {\sl 1,2,3}$. The time evolution of observables in the theory is then given by the Poisson bracket of the Hamiltonian with a local observable ${\tilde{\hat O}} (x,K)$ expressed in terms of the unconstrained independent field variables ${\tilde{\hat A}}_{i\,a}$ and $\Pitch_{j\,b}$ with positive mass-squared support 
\beq \label{79} 
\pa_{\sl 0}  {\tilde{\hat O}} (y,K)= \left\{H , {\tilde{\hat O}} (y,K) \right\}_{x_{\sl 0} = y_{\sl 0}}.
\eeq
The time evolution is compatible with the support condition Eqn.(\ref{77}). Together, Eqns.(\ref{76}) and (\ref{79}) constitute a perfectly consistent classical Hamiltonian field theory for the ${\tilde{\hat A}}_{i\,a}$-fields and their conjugates $\Pitch_{j\,b}$.

Note finally that both transformations (1) the inverse of Eqn.(\ref{75}) rotating the fields with orthogonal matrices in field space and (2) the inverse Fourier transformation back to $X$-space of fields with positive mass-squared support in $K$-space are canonical, hence allowing us equally well to start with the Hamiltonian given by Eqn.(\ref{70}) where $A_i\,^{\sl 0}$ and $\Pit_j\,^{\sl 0}$ are expressed in terms of the independent variables $A_i\,^a$ and $\Pit^j\,_b$. The positivity of the Hamiltonian is again assured by the restriction to fields whose Fourier-transformed live in the functional spaces of fields with positive mass-squared support in $K$-space, a fact which is hidden working in the original $X$-variables. The Poisson brackets though get replaced by the appropriate Dirac brackets \cite{stw1}.

\section{Inclusion of Matter Fields}
Let us finally comment on the inclusion of matter fields. The minimal coupling prescription suggests to couple matter by (1) allowing fields to live on ${\bf M}^{\sl 4}\times {\bf M}^{\sl 4}$ - adding the necessary additional inner degrees of freedom - and by (2) replacing ordinary derivatives through covariant ones $\pa_\mu\ar D_\mu$ in matter Lagrangians as usual. As this prescription involves scalars in inner space only and as the volume element $d^{\sl 4} X\, \La^{-4}$ is locally invariant, the metric $g_{\al\be}$ does not appear in minimally coupled matter actions.

Note that this prescription allows for a universal coupling of any matter field to the gauge fields of the theory treating them as scalars in inner space.

Technically no fundamentally new difficulties arise and the relevant matter terms are simply added to the formulae for both the Lagrangian and Hamiltonian gauge field theories of the group of volume-preserving diffeomorphisms of ${\bf M\/}^{\sl 4}$ \cite{stw2}.

\section{Conclusions}

In this paper we have started to explore the consequences of viewing the gravitational energy momentum $p_G^\mu$ as different by its very nature from the inertial energy-momentum $p_I^\mu$, accepting their observed numerical equality as accidential.

This view has motivated us to add new field degrees of freedom allowing to represent an inner translation group in field space in order to generate a new conserved four-vector through Noether's theorem which we interpret as gravitational energy momentum $p_G^\mu$.

Gauging this inner translation group has naturally led to the gauge field theory of the group of volume-preserving diffeomorphisms of ${\bf R\/}^{\sl 4}$ with unimodular Jacobian, at the classical level, thereby generalizing non-Abelian gauge field theories with a finite number of gauge fields. In contrast to that case, in order to gauge coordinate transformations of an inner ${\bf R\/}^{\sl 4}$ we had to introduce an uncountably infinite number of gauge fields labeled by $X^\al$, the inner coordinates of the fields on which we represent the global and local gauge groups.

This has not brought along fundamental difficulties as far as the definitions of the covariant derivative, the gauge field and the field strength operators are concerned. As the components of these operators are vectors in inner space we then introduced a flat metric $g$ on ${\bf R\/}^{\sl 4}$ in order to allow for coordinate-invariant contractions of inner space indices, making it the inner Minkowski space ${\bf M\/}^{\sl 4}$.

Potentially fundamental difficulties, however,  have arisen in the definition of other  crucial elements of the theory - such as the trace operation in the definition of the gauge field action. Tr turned out to be a potentially divergent integral over the non-compact inner ${\bf M\/}^{\sl 4}$. Accordingly we had to defined the trace operation using the scale parameter $\La$ inherent to the theory as a cut-off and have shown that the theories for different $\La$ are in fact related to each other by an additional global inner scale symmetry of the theory.

We then have proposed - with consistent quantization in view - a covariant, minimal gauge field Lagrangian. Next, we have derived the field equations and shown their independence of the inner metric $g$. Finally we have determined the conserved Noether currents and charges belonging to the inner and spacetime symmetries of the theory including the energy-momentum density of the gauge fields. 

The natural framework to consistently deal with gauge fixing, to implement the constraints and to both define a classical field theory and prepare its path integral quantization is the Hamiltonian formalism for which we have derived the theory's Hamiltonian and the corresponding Hamiltonian dynamics through choosing Cartesian coordinates with a Minkowski metric in inner space - partially fixing a gauge - and imposing on top the axial gauge condition to fully fix the gauge.

A key condition for the viability of the theory is the positivity of the Hamiltonian. A careful analysis relates $H \geq 0$ to a quite natural restriction for the support of the Fourier-transformed gauge fields being limited to the forward and backward light cones in inner $P$-space. In addition, this analysis uncovers a set of two times six unconstrained independent canonical variables obeying the usual Poisson brackets with which the theory can be formulated and which will serve as the starting point for quantization \cite{chw3}.

The result is a classical field theory formulated on flat four-dimensional Minkowski spacetime which is invariant under local ${\overline{DIFF}}\,{\bf M}^{\sl 4}$ gauge transformations and at most quartic in the fields - a perfect candidate for a renormalizable, asymptotically free quantum field theory.

In two separate papers we show that the present theory encompasses classical gravitation at the Newtonian level in a natural way \cite{chw2} and that the quantized gauge theory of volume-preserving diffeomorphisms of ${\bf M\/}^{\sl 4}$ is renormalizable and asymptotically free at one-loop \cite{chw3}.

\appendix

\section{Notations and Conventions}

Generally, ({\bf M}$^{\sl 4}$,\,$\eta$) denotes the four-dimensional Minkowski space with metric $\eta=\mbox{diag}(-1,1,1,1)$, small letters denote spacetime coordinates and parameters and capital letters denote coordinates and parameters in inner space.

Specifically, $x^\la,y^\mu,z^\nu,\dots\,$ denote Cartesian spacetime coordinates. The small Greek indices $\la,\mu,\nu,\dots$ from the middle of the Greek alphabet run over $\sl{0,1,2,3}$. They are raised and lowered with $\eta$, i.e. $x_\mu=\eta_{\mu\nu}\, x^\nu$ etc. and transform covariantly w.r.t. the Lorentz group $SO(\sl{1,3})$. Partial differentiation w.r.t to $x^\mu$ is denoted by $\pa_\mu \equiv \frac{\pa\,\,\,}{\pa x^\mu}$. Small Latin indices $i,j,k,\dots$ generally run over the three spatial coordinates $\sl{1,2,3}$ \cite{stw1}.

$X^\al, Y^\be, Z^\ga,\dots\,$ denote inner coordinates and $g_{\al\be}$ the flat metric in inner space with signature $-,+,+,+$. The metric transforms as a contravariant tensor of Rank 2 w.r.t. ${\overline{DIFF}}\,{\bf M}^{\sl 4}$. Because Riem$(g) = 0$ we can always globally choose Cartesian coordinates and the Minkowski metric $\eta$ which amounts to a partial gauge fixing to Minkowskian gauges. The small Greek indices $\al,\be,\ga,\dots$ from the beginning of the Greek alphabet run again over $\sl{0,1,2,3}$. They are raised and lowered with $g$, i.e. $x_\al=g_{\al\be}\, x^\be$ etc. and transform as vector indices w.r.t. ${\overline{DIFF}}\,{\bf M}^{\sl 4}$. Partial differentiation w.r.t to $X^\al$ is denoted by $\nabla_\al \equiv \frac{\pa\,\,\,}{\pa X^\al}$. 

The same lower and upper indices are summed unless indicated otherwise.

\end{document}